\documentclass[11pt,twoside]{article}
\usepackage{asp2014}
\usepackage{graphicx}

\aspSuppressVolSlug
\resetcounters

\begin{document}

\title{Do the constants of nature couple to strong gravitational fields?}
\author{S. P. Preval,$^1$ M. A. Barstow,$^1$, J. B. Holberg$^2$, J. D. Barrow$^3$, J. C. Berengut$^4$, J. K. Webb$^4$, D. Dougan$^4$, and J. Hu$^4$
\affil{$^1$Department of Physics and Astronomy, University of Leicester, University Road, Leicester, LE1 7RH, UK; \email{sp267@le.ac.uk}, \email{mab@le.ac.uk}}
\affil{$^2$Lunar and Planetary Laboratory, Sonnett Space Sciences Bld, University of Arizona, Tucson, AZ 85721, USA; \email{holberg@argus.lpl.arizona.edu}}
\affil{$^3$DAMPT, Centre for Mathematical Sciences, University of Cambridge, Wilberforce Road, Cambridge, CB3 0WA, UK; \email{jdb34@hermes.cam.ac.uk}}
\affil{$^4$School of Physics, University of New South Wales, Sydney, New South Wales 2052, Australia; \email{jcb@phys.unsw.edu.au}, \email{jkw@phys.unsw.edu.au}, \email{darren.dougan@hindmarsh.com.au}, \email{jiting.hu@student.unsw.edu.au}}}

\paperauthor{Simon P. Preval}{sp267@le.ac.uk}{ORCID_Or_Blank}{University of Leicester}{Department of Physics and Astronomy}{Leicester}{State/Province}{LE1 7RH}{UK}
\paperauthor{Martin A. Barstow}{mab@le.ac.uk}{ORCID_Or_Blank}{University of Leicester}{Department of Physics and Astronomy}{Leicester}{State/Province}{LE1 7RH}{UK}
\paperauthor{Jay B. Holberg}{holberg@argus.lpl.arizona.edu}{ORCID_Or_Blank}{University of Arizona}{Lunar and Planetary Laboratory}{Tucson}{State/Province}{AZ 85721}{USA}
\paperauthor{John Barrow}{jdb34@hermes.cam.ac.uk}{ORCID_Or_Blank}{University of Arizona}{DAMPT, Centre for Mathematical Sciences}{Cambridge}{State/Province}{CB3 0WA}{UK}
\paperauthor{Julian Berengut}{jcb@phys.unsw.edu.au}{ORCID_Or_Blank}{University of New South Wales}{School of Physics}{Sydney}{New South Wales}{2052}{Australia}
\paperauthor{John Webb}{jkw@phys.unsw.edu.au}{ORCID_Or_Blank}{University of New South Wales}{School of Physics}{Sydney}{New South Wales}{2052}{Australia}
\paperauthor{Darren Dougan}{darren.dougan@hindmarsh.com.au}{ORCID_Or_Blank}{University of New South Wales}{School of Physics}{Sydney}{New South Wales}{2052}{Australia}
\paperauthor{Jiting Hu}{jiting.hu@student.unsw.edu.au}{ORCID_Or_Blank}{University of New South Wales}{School of Physics}{Sydney}{New South Wales}{2052}{Australia}

\begin{abstract}
Recently, white dwarf stars have found a new use in the fundamental physics community. Many prospective theories of the fundamental interactions of Nature allow traditional constants, like the fine structure constant $\alpha$, to vary in some way. A study by \cite{berengut13a} used the Fe/Ni {\sc v} line measurements made by \cite{preval13a} from the hot DA white dwarf G191-B2B, in an attempt to detect any variation in $\alpha$. It was found that the Fe {\sc v} lines indicated an increasing alpha, whereas the Ni {\sc v} lines indicated a decreasing alpha. Possible explanations for this could be misidentification of the lines, inaccurate atomic data, or wavelength dependent distortion in the spectrum. We examine the first two cases by using a high S/N reference spectrum from the hot sdO BD+28$^{\circ}$4211 to calibrate the Fe/Ni {\sc v} atomic data. With this new data, we re-evaluate the work of \cite{berengut13a} to derive a new constraint on the variation of alpha in a gravitational field.
\end{abstract}

\section{Introduction}
Fundamental theories of everything and grand unification predict the coupling of fundamental constants to gravitational fields. In particular, variation of $\alpha$ presents itself as characteristic shifts in absorption features. The magnitude of this shift is dependent upon the transition being observed, the atomic number of the atom/ion, the ionisation stage, and the strength of the gravitational field in which the ion resides (proportional to the dimensionless potential, $GM/c^{2}r$). Heavy, highly ionised atoms are more sensitive to changes in $\alpha$. An attempt was made by \cite{berengut13a} to detect variations of $\alpha$ in the hot DA white dwarf G191-B2B ($T_{\mathrm{eff}}$=60,000K, log $g$=7.61) by measuring several features of Fe/Ni {\sc v} against their respective sensitivity indices, which is a measurement of how sensitive a transition is to a change in $\alpha$. The G191-B2B spectrum came from \cite{preval13a}, which was constructed by coadding 32 E140H Space Telescope Imaging Spectrometer (STIS) observations. The spectrum covers 1160-1645\AA, and has exceptionally high S/N, exceeding 100 in places. The authors found two conflicting values for possible $\alpha$ variation. The Fe {\sc v} lines gave $\Delta\alpha/\alpha=(4.2\pm{1.6})\times{10^{-5}}$ while the Ni {\sc v} lines gave $\Delta\alpha/\alpha=(-6.1\pm{5.8})\times{10^{-5}}$. It was suggested that the accuracy of the atomic data may have played a part in the discrepancy, however, it was also noted that there may have been a systematic effect in the wavelength calibration. One way to address both issues is to calibrate the Fe/Ni lines using the spectrum of another object with lower mass (and hence smaller gravitational field strength). BD+28$^{\circ}$4211 is a hot sdO ($T_{\mathrm{eff}}$=82,000K, log $g$=6.2), and has been observed extensively with STIS.

In this proceeding, we describe the spectroscopic survey we performed of BD+28$^{\circ}$4211, and using the newly calibrated Fe/Ni {\sc v} data, we also show that potential $\alpha$ variation can be probed down to sensitivities of $\sim{10^{-6}}$.

\section{Observations}
\begin{figure}
\begin{centering}
\includegraphics[width=100mm,clip=true,trim=0 0 0 5mm]{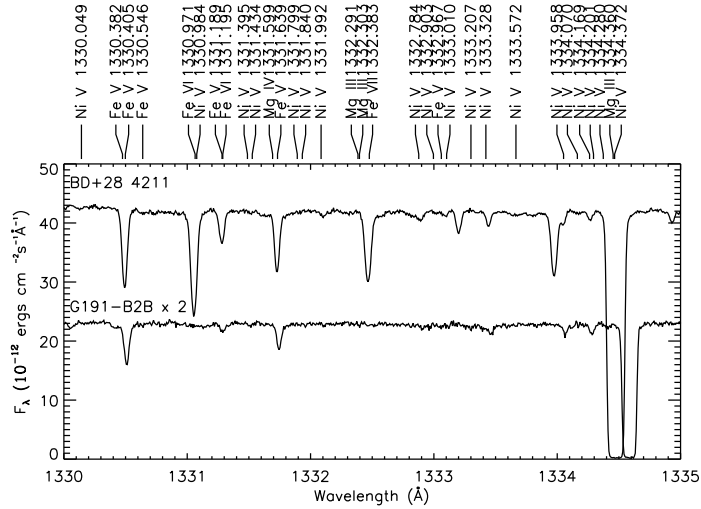}
\caption{Comparison of coadded spectra of G191-B2B and BD+28$^{\circ}$4211. The flux of G191-B2B has been multiplied by a factor of 2 for clarity. The large features near 1334.5\AA\, is interstellar C {\sc ii}.}
\label{fig:bd28g191}
\end{centering}
\end{figure}

53 E140M and 24 E140H observations from STIS were coadded to construct two spectra, one being medium resolution with $\Delta\lambda$=0.01\AA\, spanning 1160-1720\AA, and one being high resolution with $\Delta\lambda$=0.075\AA\, spanning 1315-1515\AA. The S/N of both spectra are far in excess of the coadded G191-B2B spectrum utilised by \cite{preval13a}. In Figure \ref{fig:bd28g191}, a comparison between the G191-B2B and BD+28$^{\circ}$4211 spectrum is shown.

\section{Method}
To calibrate the Fe/Ni {\sc v} data, we first measure all of the absorption features present in the coadded BD+28$^{\circ}$4211 spectra, and identify their origin (photospheric or interstellar). Note that the photospheric velocity is a combination of the gravitational redshift of the star, and the space motion. As has been shown by \cite{ayers08a} in the Deep Lamp Project, the wavelength calibration of E140M spectra is of similar quality to the E140H spectra. Where possible, we use the highest resolution spectra to measure the absorption features, so the E140M spectrum is used in the wavelength ranges 1160-1315\AA\, and 1515-1720\AA, and the E140H spectrum is used in the wavelength range 1315-1515\AA. After identifying all of the photospheric lines with velocities $v_i$, the average photospheric velocity ($v_{\mathrm{BD+28}}$) is calculated, weighted by the inverse square error $\delta{v_i}$ of the individual lines:
\begin{equation}
v_{\mathrm{BD+28}}=\frac{\sum_{i=1}^{N}\frac{v_{i}}{\delta{v_i}^2}}{\sum_{i=1}^{N}\frac{1}{\delta{v_i}^2}}
\end{equation}
The error on $v_{\mathrm{BD+28}}$ is then:
\begin{equation}
\delta{v_{\mathrm{BD+28}}}=\sqrt{\frac{1}{\sum_{i=1}^{N}\frac{1}{\delta{v_i}^2}}}
\end{equation}
Next, to assess the accuracy of the Fe/Ni {\sc v} data, we work backwards, and use the determined photospheric velocity to calculate a "laboratory wavelength", $\lambda_{\mathrm{lab}}$, based on the observed wavelength of the feature. This is calculated as:
\begin{equation}
\lambda_{\mathrm{lab}}=\frac{\lambda_{\mathrm{obs}}}{1+\frac{v_{\mathrm{BD+28}}}{c}}
\end{equation}
The calculated $\lambda_{\mathrm{lab}}$ is compared with the tabulated wavelengths \citep{kurucz11a} use to calculate the photospheric velocity. If the difference between these two values <0.001\AA, then the tabulated laboratory wavelength is regarded as good, otherwise, it is regarded as poor, and is replaced by the calculated $\lambda_{\mathrm{lab}}$. After calculating the corrected wavelengths, this data is then used with the observations made in G191-B2B to repeat the analysis of \cite{berengut13a}.

\section{Results}
\begin{figure}
\begin{centering}
\includegraphics[width=100mm]{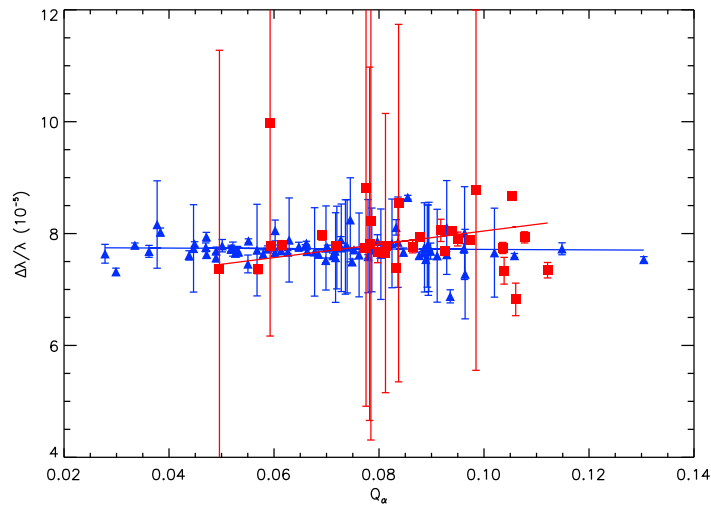}
\caption{Plot of measured redshift of Fe/Ni {\sc v} absorption features calculated with corrected wavelength data against sensitivity index $Q_{\alpha}$, where the blue triangles are Fe {\sc v}, and the red squares are Ni {\sc v}.}
\label{fig:alplot}
\end{centering}
\end{figure}

The spectroscopic survey of BD+28$^{\circ}$4211 yielded more than 1000 detections of absorption features, 671 of which were identified to originate from Fe/Ni {\sc iv-vii} transitions. 112 absorption features in the BD+28$^{\circ}$4211 spectrum were coincident with the G191-B2B spectrum, 82 of which are Fe {\sc v}, and 30 of which are Ni {\sc v}. The photospheric velocity of BD+28$^{\circ}$4211 was found to be 19.8kms$^{-1}$. Recalculating $\lambda_{\mathrm{BD+28}}$, it was found that >600 Fe/Ni features differed from the tabulated values by >1m\AA. In Figure \ref{fig:alplot}, we plot the measured redshift of the Fe/Ni {\sc v} lines using the corrected wavelengths against the sensitivity index $Q_{\alpha}$. We extracted $\alpha$ variations of $\Delta\alpha/\alpha=(3.99\pm{3.72})\times{10^{-6}}$ for the Fe lines, and $\Delta\alpha/\alpha=(-1.19\pm{0.10})\times{10^{-4}}$ for the Ni lines. While the Ni result implies a variation of $\alpha$ consistent with an $11.9\sigma$ detection, it must be stressed that the errors quoted are statistical only. The large scatter in the Ni redshifts is most likely due to misidentifications in the G191-B2B spectrum, as the Ni features are more difficult to differentiate from the Fe features due to line blending, and lower oscillator strengths in general.

\section{Conclusion}
We performed a spectroscopic survey of BD+28$^{\circ}$4211 with the aim of refining the Fe/Ni atomic data for use in constant variation studies. This method had the advantage of removing any systematics present in data take with STIS. We calculated a revised constraint on any potential $\alpha$ variation in G191-B2B as $\Delta\alpha/\alpha=(3.99\pm{3.72})\times{10^{-6}}$ for the Fe {\sc v} lines, and $\Delta\alpha/\alpha=(-1.19\pm{0.10})\times{10^{-4}}$ for the Ni {\sc v} lines. The calculated errors are statistical only, and have yet to take into account data systematics. Combined, this is consistent with a null result, however, misidentified Ni {\sc v} features may have skewed the result. Our study has shown that with high quality atomic data, potential $\alpha$ variation can be measured down to this sensitivity. Our result has also shown the effectiveness of using standard stars as stellar laboratories in order to improve the accuracy of atomic databases. 

\acknowledgements FUSE data was obtained from the Mikulski Archive for Space Telescopes (MAST). SPP and MAB acknowledge the support of the Science and Technology Facilities Council (STFC, UK). JBH acknowledges support from the NASA grant NNG056GC46G.

\end{document}